\documentclass[traditabstract,print]{aa}
\usepackage{graphicx}
\usepackage{epsfig}
\usepackage{amssymb,amsmath}
\usepackage{natbib}
\bibpunct{(}{)}{;}{a}{}{,}
\usepackage{pslatex}
\usepackage{caption}
\usepackage{longtable}
\usepackage{hyperref}

\begin{document}

\title{ Reverse Shock Emission Driven By Post-Merger Millisecond Magnetar
Winds: Effects of the Magnetization Parameter}
\author{L. D. Liu $^{1,2}$, L. J. Wang $^{3}$, Z. G. Dai $^{1,2}$\thanks{%
dzg@nju.edu.cn(ZGD)}} \institute{$^{1}$ School of Astronomy and
Space Science, Nanjing University, Nanjing
210093, China\\
$^{2}$ Key Laboratory of Modern Astronomy and Astrophysics (Nanjing
University), Ministry of Education, Nanjing 210093, China\\
$^{3}$ Key Laboratory of Space Astronomy and Technology, National
Astronomical Observatories, Chinese Academy of Sciences, Beijing
100012,
China\\
}
\authorrunning{Liu et al.}
\titlerunning{Reverse Shock Emission Driven By Post-Merger Millisecond Magnetar Winds}
\abstract{The study of short-duration gamma-ray bursts provides
growing evidence that a good fraction of double neutron star mergers
lead to the formation of stable millisecond magnetars. The launch of
Poynting flux by the millisecond magnetars could leave distinct
electromagnetic signatures that reveal the energy dissipation
processes in the magnetar wind. In previous studies (Wang \& Dai
2013b; Wang et al. 2015), we assume that the magnetar wind becomes
completely lepton-dominated so that electrons/positrons in the
magnetar wind are accelerated by a diffusive shock. However,
theoretical modeling of pulsar wind nebulae shows that in many cases
the magnetic field energy in the pulsar wind may be strong enough to
suppress diffusive shock acceleration. In this paper, we investigate
the reverse shock emission as well as the forward shock emission
with an arbitrary magnetization parameter $\sigma$ of a magnetar
wind. We find that the reverse shock emission strongly depends on
$\sigma$, and in particular, $\sigma \sim 0.3$ leads to the
strongest reverse shock emission. Future observations would be
helpful to diagnose the composition of the magnetar wind.}
 \maketitle

\section{Introduction}

\label{sec:intro}

Short-duration gamma ray bursts (SGRBs) are one of the most
mysterious phenomena in the universe that have attracted much
attention over the past decades. It is generally believed that SGRBs
result from compact binary mergers, either double neutron star
mergers or neutron star (NS) and black hole (BH) mergers
\citep{paczynski86,eichler89,barthelmy05,fox05,gehrels05}.

To definitely confirm the compact binary merger nature of SGRBs,
next-generation gravitational wave detectors are coming into online recently %
\citep{abadie10,bartos13}. To help localize the gravitational wave
sources and determine their redshifts, simultaneous multi-wavelength
electromagnetic (EM) detection is crucial. Besides SGRBs, the
neutron-rich ejecta unbound
during the mergers could power a week-long macronova or kilonova %
\citep{li98,kulkarni05,rosswog05,metzger12,berger13,tanvir13}. The
expansion
of the ejecta into its ambient medium could lead to radio afterglows %
\citep{nakar11,metzger12,piran13,rosswog13}.

With the discovery of X-ray plateau \citep{rowlinson10,rowlinson13}
and extended emission, double neutron star merger is of particular
interest
because it provides the possibility to inject additional energy %
\citep{dai98a,dai98b,zhang01} to the ejecta under the assumption
that the merger remnant is a magnetar that avoids collapse for an
astrophysically interesting period \citep{dai06,zhang13}. Based on
the assumption that a stable magnetar is formed following the double
neutron star merger, other
distinct EM signals could also be observed, e.g., X-ray transients %
\citep{zhang13}, merger-novae \citep{yu13,metzger14}, and
forward-shock afterglows \citep{gao13}.

The study of pulsar wind nebulae (PWNe) reveals that an initially
Poynting-flux dominated pulsar wind is usually converted to a wind
dominated by electron/positron pairs ($e^{\pm }$) at some large
radii. Under the
assumption that the Poynting flux becomes completely lepton dominated %
\citep{coroniti90,michel94,dai04,yu07} and therefore a reverse shock
develops within the pulsar wind, an additional EM signature could be
observed \citep{wang13b,wang15}. In this case, $e^{\pm }$ pairs in
the pulsar wind are accelerated by a diffusive shock.

In fact, the conversion efficiency of the Poynting flux to kinetic
flux cannot be 100\%. In consequence, the wind may remain strongly
magnetized after the acceleration phase. \cite{kennel84}
demonstrated that the Crab dynamics and emission can be understood
provided that the wind is weakly magnetized. There have been a
number of analytic and numerical papers on magnetization. The
existence of a reverse shock (RS) in magnetized ejecta was suggested
\citep[e.g.,][]{fan04,zhang05,giannios08,mimica09,Lyutikov11} and
the dynamics of arbitrarily magnetized ejecta-medium interaction
were studied \citep[e.g.,][]{zhang05,mimica09,Mizuno09,Lyutikov11}.
It was found
that the magnetization influences the RS emission %
\citep[e.g.,][]{mimica10,harrison13}.

In the case of magnetization the magnetic field in the shocked wind
is
determined by the shock jump conditions \citep{kennel84,fan04,zhang05,mao10}%
. The RS, arising from the interaction of the wind with the merger
ejecta
and external medium, depends on the magnetization of the magnetar wind %
\citep{fan04,zhang05,mimica09,mimica10,mao10}. The strong magnetic
fields
would suppress the RS\ in the magnetized wind %
\citep{fan04,zhang05,giannios08,mimica09,mimica10}.

The aim of this paper is to evaluate effects of the magnetization parameter $%
\sigma $ on the resulting EM emission. This paper is organized as
follows. In Section 2, we describe the model in details and in
Section 3 we show our results in different dynamical cases.
Discussion and conclusions are given in Section 4.

\section{The Model}

\label{sec:RS}

We assume that the merger product is a massive rapidly spinning
neutron star. The newborn neutron star may be close to its breakup
limit, so we take $P_{0}=1$ ms as the initial spin period. Due to
rapidly differential rotation of the newborn neutron star, the onset
of magneto-rotational instability could amplify the magnetic field
of such neutron star to magnetar levels
\citep{duncan92,magnetar-sim}. Hence, a proto-magnetar may have a
magnetic field of $\sim 10^{14}-10^{15}$ G. Its spin-down luminosity
can be expressed by
\begin{eqnarray}
L_{\text{sd}}=L_{\text{sd,0}}\left( 1+\frac{t}{T_{\text{sd}}}\right)
^{-2}, \label{Spin-Down Luminosity}
\end{eqnarray}%
where the magnetar initial spin-down luminosity
$L_{\text{sd,0}}=1\times
10^{49}$ erg s$^{-1}B_{\text{p,15}}^{2}R_{\text{s},6}^{6}P_{0,-3}^{-4},$ $B_{%
\text{p,15}}=B_{\text{p}}/10^{15}$ G is the polar-cap dipole
magnetic field and $R_{\text{s,6}}=R_{\text{s}}/10^{6}$ cm is the
stellar radius. Here we take the conventional usage $Q=10^{n}Q_{n}$.
The characteristic spin-down timescale of the millisecond magnetar
is given by
\begin{eqnarray}
T_{\text{sd}}=2\times 10^{3}\text{ s }I_{\text{s,}45}B_{p,15}^{-2}R_{\text{s}%
,6}^{-6}P_{0,-3}^{2},  \label{Spin-Down timescale}
\end{eqnarray}%
where $I_{\text{s,45}}=I_{\text{s}}/10^{45}$ g cm$^{2}$ is the
magnetar moment of inertia. For a millisecond massive magnetar, the
typical values of $I_{\text{s,45}},R_{\text{s,6}}$ and\ $P_{0,-3}$
are all close to unity.

Recently numerical simulations show that the ejecta from neutron
star binary\ mergers has a typical velocity $v_{\text{ej}}=0.1$--
$0.3c$ and a
typical mass $M_{\text{\textrm{ej}}}=10^{-4}$--$10^{-2}M_{\odot }$ %
\citep{rezzolla10,hotokezaka13,rosswog13}. The release of
millisecond magnetar rotational energy could launch a
Poynting-flux-dominated outflow
whose luminosity is determined by equation $\left( \ref{Spin-Down Luminosity}%
\right) $. This outflow would quickly catch up and interact with the
ejecta at a radius $R_{\text{ca}}\sim
v_{\text{ej}}t_{\text{delay}}=6\times 10^{10}$
cm$\left( v_{\text{ej}}/0.2c\right) t_{\text{delay,1}}$, where $t_{\text{%
delay}}=10t_{\text{delay,1}}$ s is the\ delay time between the\
merger and the launch of Poynting-flux-dominated outflow
\citep{metzger11}. A forward shock would be driven into the ejecta,
because the magnetar wind could push the ejecta to a relativistic
speed. The forward shock would cross the ejecta in a short timescale
\citep{gao13} and then the forward shock propagates into ambient
medium.

Since the fluctuating component of the magnetic field in the outflow
can be dissipated by magnetic reconnection, the Poynting flux is
converted to the kinetic energy of the outflow, and the magnetar
wind eventually becomes leptonic-matter-dominated
\citep{dai04,coroniti90,michel94}. In this
scenario, a reverse shock develops and produces additional radiation signals %
\citep{dai04,yu07,mao10,wang13b,wang15}. In some cases, however, the
magnetar wind may remain rather strongly magnetized at large radii.
The magnetar wind in these cases is more likely to be a mixture of
Poynting flux and ultra-relativistic $e^{\pm }$ pairs. To quantify
the magnetic field component in the outflow, the magnetization
parameter $\sigma $, i.e. the
ratio of the Poynting flux $F_{\text{P }}$ to the matter flux $F_{\text{m}}$%
, is introduced
\begin{eqnarray}
\sigma \equiv \frac{F_{\text{P}}}{F_{\text{m}}}=\frac{B^{2}}{4\pi
\Gamma \rho c^{2}}=\frac{B^{\prime 2}}{4\pi \rho ^{\prime }c^{2}},
\label{sigma}
\end{eqnarray}%
where $B$ and $\rho $ are the magnetic field strength and matter
density in the observer frame, and $B^{\prime }$ and $\rho ^{\prime
}$ are the corresponding quantities in the comoving frame of the
outflow.

If the magnetar wind is leptonic-matter-dominated, its bulk Lorentz factor $%
\Gamma _{\text{w}}$ is typically taken to be in the range of $10^{4}$ to $%
10^{7}$ \citep{atoyan99,michel99}. In principle, the value of $\Gamma _{%
\text{w}}$ depends on the magnetization parameter $\sigma $. For
simplicity, we here adopt $\Gamma _{\text{w}}\sim 10^{4}$ as a
fiducial value.

The basic physical picture of our model is schematized in Figure 1 of \cite%
{gao13} and Figure 1 of \cite{wang15}. The interaction of the
magnetar wind with the ejecta and ambient medium would give rise to
a relativistic wind bubble \citep{dai04,yu07,mao10,wang13b,wang15}.
Two shocks are formed: a reverse shock that propagates into the cold
magnetar wind and a forward shock propagates into the ambient
medium. Thus, the relativistic wind bubble includes four different
regions separated by a contact discontinuity surface and two shocks.
The four regions are\ (1) the unshocked ambient medium, (2) the
forward-shocked ambient medium, (3) the reverse-shocked magnetar
wind, and (4) the unshocked cold wind. Regions 2 and 3 are divided
by a contact discontinuity surface. The shocked regions produce
synchrotron emission. Hereafter, we denote the quantities of Region
$i$ as follows: $n_{i}^{\prime },B_{i}^{\prime },e_{i}^{\prime }$
and $P_{i}^{\prime }$ are particle number density, magnetic field
strength, energy density and pressure, where the primes refer to the
quantities in the comoving frame, and subscript $i$ represents
Region 1-4.

In general, the ambient medium is unmagnetized so that $\sigma
_{1}=0$. The magnetization degree of the magnetar wind could be
quantified by the magnetization parameter
\begin{eqnarray}
\sigma \equiv \sigma _{4}=\frac{B_{4}^{\prime 2}}{4\pi n_{4}^{\prime
}m_{e}c^{2}},
\end{eqnarray}%
Assuming that the magnetar wind is isotropic, the comoving electron
density
of the unshocked cold wind is given by%
\begin{eqnarray}
n_{4}^{\prime }=\frac{L_{0}}{4\pi R^{2}\Gamma
_{4}^{2}m_{e}c^{3}\left( 1+\sigma \right) },
\end{eqnarray}%
where $L_{0}=\xi L_{\text{sd,0}}$ with $\xi $ denoting the fraction
of the spin-down power injected into the shocks, $R$ is the radius
of the wind bubble, $\Gamma _{4}$ is the Lorentz factor of Region 4
and in our calculation we take $\Gamma _{4}=\Gamma _{\text{w}}$.
Under ideal MHD conditions, due to the induction equation the
magnetic field in the wind is
dominated by its toroidal component \citep{fan04,zhang05,mao10}. Both $%
n_{4}^{\prime }$ and $B_{4}^{\prime 2}$\ would decrease as $R^{-2},$
as a result the magnetization parameter $\sigma $\ can be regarded
as a constant.

The connection of the two sides of shocks can be described by the
shock jump conditions \citep{coroniti90,zhang05,mao10}
\begin{eqnarray}
\frac{e_{2}^{\prime }}{n_{2}^{\prime }m_{p}c^{2}} &=&\Gamma _{2}-1,
\notag
\\
\frac{n_{2}^{\prime }}{n_{1}} &=&4\Gamma _{2}+3,  \notag \\
\frac{e_{3}^{\prime }}{n_{3}^{\prime }m_{e}c^{2}} &=&\left( \Gamma
_{34}-1\right) f_{a}, \\
\frac{n_{3}^{\prime }}{n_{4}^{\prime }} &=&\left( 4\Gamma
_{34}+3\right)
f_{b},  \notag \\
\frac{B_{3}^{\prime }}{B_{4}^{\prime }} &=&\left( 4\Gamma
_{34}+3\right) f_{b},\text{ \ for }\sigma >0,  \notag
\end{eqnarray}%
where $m_{p}$ is the proton rest mass, $\Gamma _{34}\approx \frac{1}{2}%
\left( \Gamma _{3}/\Gamma _{4}+\Gamma _{4}/\Gamma _{3}\right) $ is
the bulk Lorentz factor of Region 3 measured in the comoving frame
of Region 4, $f_{a} $ and $f_{b}$\ are the correction factors for
$e_{3}^{\prime }/n_{3}^{\prime
}m_{e}c^{2}$ and $n_{3}^{\prime }/n_{4}^{\prime }$ with respect to the $%
\sigma =0$\ case. When $\Gamma _{34}\gg 1$, $f_{a}$ and $f_{b}$ can
be expressed as \citep{zhang05,mao10} \
\begin{eqnarray}
f_{a} &\approx &1-\frac{\sigma }{2\left[ u_{\text{3s}}^{2}+u_{\text{3s}%
}\left( u_{\text{3s}}^{2}+1\right) ^{1/2}\right] },  \label{fa} \\
f_{b} &\approx &\frac{1}{4}\left[ 1+\left( 1+\frac{1}{u_{\text{3s}}^{2}}%
\right) ^{1/2}\right] ,  \label{fb}
\end{eqnarray}%
where $f_{a}\rightarrow 1$ and $f_{b}\rightarrow 1$ when $\sigma
\rightarrow 0$, and $u_{\text{3s}}=\Gamma _{3}\beta _{3}$ is the
four-velocity of Region 3 measured in the rest frame comoving with
the reverse shock. When $\Gamma
_{34}\gg 1,$ $u_{\text{3s}}$ can be easily\ solved analytically from%
\citep{fan04,zhang05}
\begin{equation*}
8\left( \sigma +1\right) u_{\text{3s}}^{4}-\left( 8\sigma
^{2}+10\sigma +1\right) u_{\text{3s}}^{2}+\sigma ^{2}=0.
\end{equation*}%
Equal speed and pressure equilibrium across the contact
discontinuity surface give
\begin{eqnarray}
\Gamma _{2} &=&\Gamma _{3}, \\
P_{3}^{\prime } &=&P_{\text{th,3}}^{\prime }+P_{\text{B,3}}^{\prime
}=P_{2}^{\prime },  \label{pressure-condition}
\end{eqnarray}%
where $P_{\text{th,3}}^{\prime }$ is the thermal pressure and $P_{\text{B,3}%
}^{\prime }$\ is the\ magnetic pressure of Region 3. According to
the jump conditions we can obtain $P_{\text{th,3}}^{\prime
}=\frac{1}{3}e_{3}^{\prime }\approx \frac{4}{3}\Gamma
_{34}^{2}n_{4}^{\prime }m_{e}c^{2}f_{a}f_{b}$ and
$P_{\text{B,3}}^{\prime }=\frac{B_{3}^{\prime 2}}{8\pi }\approx
8\Gamma _{34}^{2}n_{4}^{\prime }m_{e}c^{2}\sigma f_{b}^{2}.$ The
total kinetic energy of Region 2 is approximated by
\begin{eqnarray}
E_{\text{k,2}}=\left( \Gamma _{2}-1\right) \left( M_{\text{ej}}+M_{\text{sw}%
}\right) c^{2}+\Gamma _{2}\left( \Gamma _{2}-1\right)
M_{\text{sw}}c^{2},
\end{eqnarray}%
where $M_{\text{sw}}$ is the swept-up medium mass. Due to energy
conservation, any increase of $E_{\text{k,2}}$\ should be equal to
the work
done by Region 3,%
\begin{eqnarray}
dE_{\text{k,2}}=\delta W=4\pi R^{2}P_{3}^{\prime }dR.
\end{eqnarray}%
Then we can obtain
\begin{eqnarray}
\frac{d\Gamma _{2}}{dR}=\frac{4\pi R^{2}}{M_{\text{ej}}+2\Gamma
_{2}M_{\text{ sw}}}\left[ \frac{P_{3}^{\prime }}{c^{2}}-\left(
\Gamma _{2}^{2}-1\right) n_{1}m_{p}\right] .  \label{eq:gamma2-R}
\end{eqnarray}

The evolutions of the swept-up mass $M_{\text{sw}}$ and the radius
$R$ of the shock \ can be described as \citep{huang99}
\begin{eqnarray}
\frac{dM_{\text{sw}}}{dR} &=&4\pi R^{2}n_{1}m_{p}, \\
\frac{dR}{dt} &=&\frac{\beta c}{1-\beta }.  \label{eq:R-dot}
\end{eqnarray}

The energy injected by the magnetar is deposited both in the forward
shock and the reverse shock. In the non-magnetized case, the energy
contained in the forward shock is
\begin{eqnarray}
E_{\text{fs}}\approx \left( \Gamma ^{2}-1\right) M_{\text{sw}}c^{2},
\end{eqnarray}%
which is determined by the shock jump conditions. The energy
deposited in
reverse shock $E_{\text{rs}}$\ is comparable to $E_{\text{fs}}$\ %
\citep{blandford76,wang13b,wang15}. In an arbitrarily magnetized
wind, the total energy of the wind is in kinetic (electron-positron
pairs) and magnetic forms. Their fractions are $\frac{1}{1+\sigma }$
and $\frac{\sigma }{1+\sigma }$, respectively. The pairs in the
magnetar wind will be
reverse-shocked upon interaction with the ambient medium to gain an energy $%
\frac{1}{\sigma +1}\left( \Gamma ^{2}-1\right) M_{\text{sw}}c^{2}$.
Consequently, the total energy contained in the blast wave is $\left( \frac{1%
}{\sigma +1}+1\right) \left( \Gamma ^{2}-1\right)
M_{\text{sw}}c^{2}.$ The dynamics of the blast wave could be
determined by
\begin{eqnarray}
L_{0}\min \left( t,T_{\text{sd}}\right) =\left( \Gamma -\Gamma _{\text{ej,0}%
}\right) M_{\text{ej}}c^{2}+\left( \frac{\sigma +2}{\sigma
+1}\right) \left( \Gamma ^{2}-1\right) M_{\text{sw}}c^{2},
\label{dynamic from wang13b}
\end{eqnarray}%
where $\Gamma $ is the Lorentz factor of the ejecta with initial
Lorentz factor $\Gamma _{\text{ej,0}}.$\quad The dynamic equation in
\cite{wang13b} does not take the magnetization into account.
\cite{wang13b} adopted the
factor of 2 in the second term on the right hand side. The factor $\frac{%
\sigma +2}{\sigma +1}$ is reduced to the non-magnetized case when
the magnetization parameter $\sigma $ approaches zero. For the
Poynting flux-dominated magnetar wind, $\sigma $ is high enough so
that there is no reverse shock into the wind. The factor is reduced
to the \cite{gao13} situation. Therefore, this factor can be used in
more general cases.

Initially, $\left( \Gamma -\Gamma _{\text{ej,0}}\right) M_{\text{ej}%
}c^{2}\gg \left( \frac{\sigma +2}{\sigma +1}\right) \left( \Gamma
^{2}-1\right) M_{\text{sw}}c^{2}$, so that the ejecta would be
accelerated linearly with time until $t=\min \left( T_{\text{
sd}},T_{\text{dec}}\right)
$. The deceleration timescale $T_{\text{dec}}$ is defined by the condition $%
\left( \Gamma -\Gamma _{\text{ej,0}}\right) M_{\text{ej}}c^{2}=\left( \frac{%
\sigma +2}{\sigma +1}\right) \left( \Gamma ^{2}-1\right) M_{\text{sw}}c^{2}$%
. The dynamical evolution of the ejecta depends strongly on the mass
of the ejecta $M_{\text{ej}}$. By setting $T_{\text{sd}}\sim
T_{\text{dec}}$, we can obtain a critical ejecta mass
\citep{gao13,wang13b,wang15}
\begin{eqnarray}
M_{\text{ej,c}}\sim 10^{-3}M_{\odot
}n^{1/8}I_{45}^{5/4}L_{0,47}^{-3/8}P_{0,-3}^{-5/2}\xi ^{5/4}\left( \frac{%
\sigma +2}{\sigma +1}\right) ^{1/8}.  \label{critical ejecta mass}
\end{eqnarray}

It is this critical mass that divides the dynamics of the ejecta
into three
cases \citep{gao13,wang13b,wang15}: Case I for $M_{\text{ej}}<M_{\text{ej,c}%
},$ which is equivalent to $T_{\text{dec}}<T_{\text{sd}}$, Case II for $T_{%
\text{dec}}=T_{\text{sd}}$, and Case III for
$T_{\text{dec}}>T_{\text{sd}}$. With the propagation of the shocks,
the bulk kinetic energy of the ejecta and the rotational energy of
the magnetar wind would be gradually transformed into internal
energy of the shocked matter. Both the forward shock and the reverse
shock could heat the cold materials, and particles could be
accelerated by these relativistic shocks. We consider synchrotron
radiation of the accelerated electrons in the shocked regions. It is
similar to the standard model of gamma-ray burst afterglows, whose
spectrum includes four power-law segments separated by three break
frequencies, i.e., the
self-absorption frequency $\nu _{a}$, the characteristic frequency $\nu _{m}$%
, and the cooling frequency $\nu _{c}$, with peak flux $F_{\nu ,\max }$ %
\citep{sari98}. We obtain the self-absorption frequency $\nu _{a}$
according to \cite{wu03}.

The forward shock emission is calculated quantitatively in a way
similar to \cite{gao13}. For the reverse-shocked wind, the energy
density and the\ magnetic field can be determined by the shock jump
conditions for an
arbitrary magnetization parameter $\sigma $. The minimum Lorentz factor of $%
e^{\pm }$ pairs in the shocked wind is
\begin{eqnarray}
\gamma _{m}=\frac{p-2}{p-1}\frac{e_{3}^{\prime }}{n_{3}^{\prime
}m_{e}c^{2}}, \label{eq:gamma-m}
\end{eqnarray}%
where $p$ is the power-law index of the electron energy distribution
in Region 3. To find the maximum Lorentz factor of electrons, two
timescales
are considered here. The first is the particle acceleration timescale%
\begin{eqnarray}
t_{\mathrm{acc}}=\frac{\gamma _{e}m_{e}c}{q_{e}B^{\prime }}.
\end{eqnarray}%
The other one is the electron cooling time%
\begin{eqnarray}
t_{\mathrm{cool}}=\frac{\gamma
_{e}m_{e}c^{2}}{P_{\mathrm{rad}}^{\prime }},
\end{eqnarray}%
where the radiation power is given by%
\begin{eqnarray}
P_{\mathrm{rad}}^{\prime }=\frac{4}{3}\sigma _{T}c\gamma _{e}^{2}\frac{%
B^{\prime 2}}{8\pi }.
\end{eqnarray}%
Identifying these two timescales then we have the maximum Lorentz factor%
\begin{eqnarray}
\gamma _{\max }\approx \left( \frac{6\pi q_{e}}{\sigma _{T}B^{\prime }}%
\right) ^{1/2}.
\end{eqnarray}

The cooling Lorentz factor of the electrons accelerated by the
reverse shock is given by
\begin{eqnarray}
\gamma _{c}=\frac{6\pi m_{e}c}{\sigma _{T}\Gamma _{3}B_{3}^{\prime
2}t},
\end{eqnarray}%
where $\sigma _{T}$ is the Thompson cross section. Due to the
braking caused by the ejecta, at the beginning the magnetar wind can
only drive the ejecta in a small radius, leading to a high energy
density and a strong magnetic field in Region 3. Consequently, the
cooling of $e^{\pm }$ pairs is so quick that the cooling Lorentz
factor $\gamma _{c}\approx 1,$ and this Lorentz
factor dose not deviate significantly from unity until after $T_{\text{ct}}$ %
\citep{wang13b}. The fast cooling feature of the reverse shock before $T_{%
\text{ct}}$ seems to be a concern about the validity of Equation $\left( \ref%
{eq:gamma-m}\right) $ because this equation is only correct if the
upper energy cutoff $\gamma _{\max }$ is much larger than $\gamma
_{m}$. It is found that in our calculation $\gamma _{\max }\gg
\gamma _{m}$\ is always satisfied. Thus our results are almost
independent of $\gamma _{\max }$.
This is true because here we have three timescales: the dynamical timescale $%
t$, the cooling timescale $t_{\mathrm{cool}}$, and the acceleration
timescale $t_{\mathrm{acc}}$. When $t<T_{\text{ct}}$, we have $t_{\mathrm{acc%
}}\ll t_{\mathrm{cool}}\ll t$. That is, the electrons can be
accelerated to high energy before they are cooled by synchrotron
radiation.

At the spin down time $T_{\text{sd}}$, the central engine would turn
off and the unshocked wind (Region 4) would disappear and thereafter
Region 3 begins to spread linearly. Assume that the width of the
reverse-shocked wind in the comoving frame $\Delta
_{\text{RS}}\propto R.$ Before the spin down time of the magnetar,
the total number of swept-up $e^{\pm }$ pairs in the shocked wind is
$N_{\text{3,e}}=\frac{L_{0}t}{\Gamma _{4}m_{e}c^{2}\left( 1+\sigma
\right) }.$ After $T_{\text{sd}}$, Region 4 does not exist and there
are no new accelerated electrons. Hence, the total number of
particles in the emission region is a constant. Due to spread of the
width of Region 3, the
number density and energy density would decrease. When $t>T_{\text{sd}},$%
\begin{eqnarray}
n_{3}^{\prime }=n_{3\text{,sd}}^{\prime }\frac{\Gamma
_{\text{sd}}}{\Gamma _{3}}\left( \frac{R_{\text{sd}}}{R}\right)
^{3},
\end{eqnarray}%
\qquad where $n_{3\text{,sd}}^{\prime },$ $\Gamma _{\text{sd}}$ and $R_{%
\text{sd}}$ are the number density, Lorentz factor and radius at the
spin down time $T_{\text{sd}},$\ respectively. Furthermore, since
the $e^{\pm }$ pairs are in the slow cooling regime after
$T_{\text{sd}}$, $\gamma _{m}$ should keep a constant.

\section{Results}

For a quantitative analysis, we show analytical results of the
reverse shock based on the dynamic equation $\left( \ref{dynamic
from wang13b}\right) $ for the three cases mentioned above.
Numerical results including the dynamics, evolution of typical
frequencies of the reverse shock, and light curves of emission from
the shocked regions, obtained by the fourth-order
Runge-Kutta method to the dynamic evolution Equations $\left( \ref%
{eq:gamma2-R}\right) -\left( \ref{eq:R-dot}\right) $, are presented
in Figures 1 to 3.

Case I: $M_{\text{ej}}<M_{\text{ej,c}}$, or
$T_{\text{dec}}<T_{\text{sd}}$.
According to equation $\left( \ref{critical ejecta mass}\right) $ $M_{\text{%
ej,c}}\propto L_{0}^{-3/8},$ so we consider a small $L_{0}$ and a small $M_{%
\text{ej}}$ to satisfy $M_{\text{ej}}<M_{\text{ej,c}}$. We take the\
same
parameters as \cite{gao13} $L_{0}=10^{47}$ erg s$^{-1}$ and $M_{\text{ej}%
}=10^{-4}M_{\odot }$. In order to describe the dynamics in this
case, we need list the characteristic timescales of the blast wave
dynamics and the Lorentz factor at the deceleration time
\citep{gao13,wang13b,wang15}
\begin{eqnarray}
T_{\text{N1}} &=&2.07\times 10^{-2}\text{ days }L_{0,47}^{-1}M_{\text{ej,}%
-4},  \notag \\
T_{\text{dec}} &=&0.29\text{ days }L_{0,47}^{-7/10}M_{\text{ej,}%
-4}^{4/5}n^{-1/10}\left( \frac{\sigma +2}{\sigma +1}\right) ^{7/10},
\notag
\\
T_{\text{sd}} &=&2.3\text{ days }L_{0,47}^{-1}, \\
T_{\text{N2}} &=&28.58\text{ days }L_{0,47}^{1/3}T_{\text{sd}%
,5}^{1/3}n^{-1/3},  \notag \\
\Gamma _{\text{dec}} &=&7.17L_{0,47}^{3/10}M_{\text{ej,}-4}^{-1/5}n^{-1/10}%
\left( \frac{\sigma +2}{\sigma +1}\right) ^{-3/10}+1,  \notag
\end{eqnarray}%
where $T_{\text{N1}}$ is the timescale of the blast wave evolving
from the non-relativistic to relativistic phase due to energy
injection, which is defined by $\gamma -1=1,$ and $T_{\text{N2}}$ is
the timescale of the blast wave evolving from the relativistic to
non-relativistic phase due to deceleration by the ambient medium.

Although the dynamical evolution is almost insensitive to the
magnetization parameter $\sigma $, the magnetic field in Region 3 is
predominantly due to the compression of the upstream magnetic field
so that the $\sigma $ of the unshocked wind would determine the
magnetic field strength of the reverse shock. In consequence, the
$\sigma $ value could affect the characteristic timescales and the
break frequencies of the synchrotron radiation.
\begin{eqnarray}
T_{\text{ct}} &=&3.2\times 10^{-3}\text{ days }L_{0,49}^{-2/3}M_{\text{ej,}%
-4}^{5/6}f_{b}^{1/3}\left( \frac{\sigma }{1+\sigma }\right) ^{1/6},
\notag
\\
T_{ac} &=&4.1\times 10^{-3}\text{ days }%
c_{1}^{1/23}L_{0,49}^{-31/46}n_{1}^{1/23} \\
&&\times M_{\text{ej},-4}^{20/23}f_{b}^{9/23}\left( \frac{\sigma }{1+\sigma }%
\right) ^{9/46},  \notag \\
T_{mc} &=&6.6\times 10^{-3}\text{ days }L_{0,49}^{-5/7}  \notag \\
&&\times M_{\text{ej,}-4}^{6/7}\varepsilon _{e}^{1/7}\Gamma
_{4,4}^{1/7}f_{b}^{2/7}\left( \frac{\sigma }{1+\sigma }\right)
^{1/7}, \notag
\end{eqnarray}%
where $T_{ac}$ and $T_{mc}$\ are two critical times, when the
cooling frequency $\nu _{c}$ crosses the self-absorption frequency
$\nu _{a}$ and the characteristic frequency $\nu _{m}$,
respectively.

The break frequencies and peak flux of Region 3 at $T_{\text{dec
}}$are derived as
\begin{eqnarray}
\nu _{a,\text{dec}} &=&2.5\times 10^{9}\text{ Hz }%
c_{1}^{3/5}L_{0,47}^{33/50}M_{\text{ej,}-4}^{-6/25}\Gamma
_{4,4}^{-1}n^{19/50}f_{a}^{-1}f_{b}^{2/5}  \notag \\
&&\times \left( \frac{p-1}{p-2}\right) \left( \frac{\sigma }{1+\sigma }%
\right) ^{1/5}\left( \frac{\sigma +1}{\sigma +2}\right) ^{67/200},  \notag \\
\nu _{m,\text{dec}} &=&3.14\times 10^{13}\text{ Hz }\Gamma
_{4,4}^{2}n^{1/2}f_{a}^{2}f_{b}\left( \frac{p-2}{p-1}\right) ^{2}  \notag \\
&&\times \sqrt{\frac{\sigma }{\left( 1+\sigma \right) ^{5}}}\left( \frac{%
\sigma +1}{\sigma +2}\right) ^{1/8},  \notag \\
\nu _{c,\text{dec}} &=&1.1\times 10^{13}\text{ Hz }L_{0,47}^{1/5}M_{\text{ej,%
}-4}^{-4/5}n^{-9/10}f_{b}^{-3}  \notag \\
&&\times \left( \frac{1+\sigma }{\sigma }\right) ^{3/2}\left( \frac{\sigma +1%
}{\sigma +2}\right) ^{23/40}, \\
F_{\nu ,\max ,\text{dec}} &=&2.2\times 10^{4}\text{ mJy }L_{0,47}^{9/10}M_{%
\text{ej,}-4}^{2/5}\Gamma _{4,4}^{-1}n^{1/5}D_{27}^{-2}f_{b}  \notag \\
&&\times \frac{\sqrt{\sigma }}{\left( 1+\sigma \right) ^{3/2}}\left( \frac{%
\sigma +2}{\sigma +1}\right) ^{9/40},  \notag
\end{eqnarray}%
where $c_{1}=\frac{16\times 2^{2/3}}{3\Gamma \left( 1/3\right) }\frac{p+2}{%
3p+2}$ \citep{wu03}\ and $D_{27}$ is the luminosity distance to the
source in units of 10$^{27}$ cm.

Case II: $M_{\text{ej}}=M_{\text{ej,c}}$, or
$T_{\text{dec}}=T_{\text{sd}}$. We adopt $L_{0}=10^{49}$ erg
s$^{-1}$ and $M_{\text{ej}}=10^{-4}M_{\odot },$ which satisfies
$T_{\text{dec}}=T_{\text{sd}}$. The characteristic timescales of the
blast wave\ dynamics\ and the Lorentz factor at the
spin-down time \citep{gao13,wang13b,wang15}%
\begin{eqnarray}
T_{\text{N1}} &=&2.07\times 10^{-4}\text{ days }L_{0,49}^{-1}M_{\text{ej}%
,-4},  \notag \\
T_{\text{dec}} &=&T_{\text{sd}}=2.3\times 10^{-2}\text{ days
}L_{0,49}^{-1},
\\
T_{\text{N2}} &=&28.58\text{ days }L_{0,49}^{1/3}T_{\text{sd}%
,3}^{1/3}n^{-1/3},  \notag \\
\Gamma _{\text{dec}}
&=&55.9L_{0,49}T_{\text{sd,3}}M_{\text{ej},-4}^{-1}+1,
\notag \\
T_{\text{ct}} &=&3.2\times 10^{-3}\text{ days }L_{0,49}^{-2/3}M_{\text{ej,}%
-4}^{5/6}f_{b}^{1/3}\left( \frac{\sigma }{1+\sigma }\right) ^{1/6},
\notag
\\
T_{ac} &=&4.1\times 10^{-3}\text{ days }%
c_{1}^{1/23}L_{0,49}^{-31/46}n_{1}^{1/23}M_{\text{ej,}%
-4}^{20/23}f_{b}^{9/23}\left( \frac{\sigma }{1+\sigma }\right)
^{9/46},
\notag \\
T_{mc} &=&6.6\times 10^{-3}\text{ days }L_{0,49}^{-5/7}M_{\text{ej}%
,-4}^{6/7}\Gamma _{4,4}^{1/7}f_{a}^{1/7}f_{b}^{2/7}\left( \frac{\sigma }{%
1+\sigma }\right) ^{1/7},  \notag
\end{eqnarray}

The characteristic frequencies of the\ reverse shock and observed
peak flux in Case II are
\begin{eqnarray}
\nu _{a,\text{sd}} &=&1.9\times 10^{11}\text{ Hz }%
c_{1}^{3/5}L_{0,49}^{11/5}M_{\text{ej,-4}}^{-2}\Gamma
_{4,4}n^{19/50}  \notag
\\
&&\times f_{a}^{-1}f_{b}^{2/5}\left( \frac{\sigma }{1+\sigma
}\right) ^{1/5},
\notag \\
\nu _{m,\text{sd}} &=&2.8\times 10^{12}\text{ Hz }L_{0,49}^{-7/2}T_{\text{%
sd,3}}^{-5}M_{\text{ej,-4}}^{4}\Gamma _{4,4}^{2}n^{1/2}  \notag \\
&&\times \left( \frac{p-2}{p-1}\right)
^{2}f_{a}^{2}f_{b}\sqrt{\frac{\sigma
}{\left( 1+\sigma \right) ^{5}}}, \\
\nu _{c,\text{sd}} &=&5.5\times 10^{15}\text{ Hz }L_{0,49}^{13/2}T_{\text{%
sd,3}}^{9}M_{\text{ej,-4}}^{8}n^{-9/10}  \notag \\
&&\times f_{b}^{-3}\left( \frac{1+\sigma }{\sigma }\right) ^{3/2},  \notag \\
F_{\nu ,\max ,\text{sd}} &=&3.6\times 10^{5}\text{ mJy }L_{0,49}^{-1/2}T_{%
\text{sd,3}}^{-2}M_{\text{ej,-4}}^{2}\Gamma _{4,4}^{-1}D_{27}^{-2}  \notag \\
&&\times f_{b}\frac{\sqrt{\sigma }}{\left( 1+\sigma \right) ^{3/2}}.
\notag
\end{eqnarray}

Case III: $M_{\text{ej,c}}<M_{\text{ej}}$, or $T_{\text{sd}}<T_{\text{dec}}$%
. We take $L_{0}=10^{49}$ erg s$^{-1}$ and
$M_{\text{ej}}=10^{-3}M_{\odot }.$
Similar to Case I and Case II, we have \citep{gao13,wang13b,wang15}%
\begin{eqnarray}
T_{\text{N1}} &=&2.07\times 10^{-3}\text{ days }L_{0,49}^{-1}M_{\text{ej}%
,-3},  \notag \\
T_{\text{sd}} &=&2.3\times 10^{-2}\text{ days }L_{0,49}^{-1},  \notag \\
T_{\text{dec}} &=&5.76\text{ days }L_{0,49}^{-7/3}T_{\text{sd,3}}^{-7/3}M_{%
\text{ej,-3}}^{8/3}n^{-1/3}\left( \frac{\sigma +2}{\sigma +1}\right)
^{7/3},
\\
T_{\text{N2}} &=&28.58\text{ days }L_{0,49}^{1/3}T_{\text{sd,}%
3}^{1/3}n^{-1/3}\left( \frac{\sigma +2}{\sigma +1}\right) ^{-1/3},  \notag \\
\Gamma _{\text{dec}} &=&5.59L_{0,49}T_{\text{sd,3}}M_{\text{ej,}%
-4}^{-1}\left( \frac{\sigma +2}{\sigma +1}\right) ^{-1}+1,  \notag
\end{eqnarray}%
And the break frequencies and peak flux of Region 3 at $T_{\text{dec
}}$are derived as
\begin{eqnarray}
\nu _{a,\text{sd}} &=&1.7\times 10^{9}\text{ Hz }%
c_{1}^{3/5}L_{0,49}^{11/5}T_{\text{sd,3}}^{11/5}M_{\text{ej,}-3}^{-2}\Gamma
_{4,4}n^{3/5}  \notag \\
&&\times f_{a}^{-1}f_{b}^{2/5}\left( \frac{p-1}{p-2}\right) \left( \frac{%
\sigma }{1+\sigma }\right) ^{1/5},  \notag \\
\nu _{m,\text{sd}} &=&2.2\times 10^{16}\text{ Hz }L_{0,49}^{-7/2}T_{\text{%
sd,3}}^{-5}M_{\text{ej,}-3}^{4}\Gamma _{4,4}^{2}n^{1/2}  \notag \\
&&\times f_{a}^{2}f_{b}\left( \frac{p-2}{p-1}\right)
^{2}\sqrt{\frac{\sigma
}{\left( 1+\sigma \right) ^{5}}}, \\
\nu _{c,\text{sd}} &=&2.7\times 10^{10}\text{ Hz }L_{0,49}^{-3/2}T_{\text{%
sd,3}}^{-3}M_{\text{ej,}-3}^{2}f_{b}\left( \frac{\sigma }{1+\sigma
}\right)
^{1/2},  \notag \\
F_{\nu ,\max ,\text{sd}} &=&3.5\times 10^{7}\text{ mJy }L_{0,49}^{-1/2}T_{%
\text{sd,3}}^{-2}M_{\text{ej,}-3}^{2}\Gamma
_{4,4}^{-1}n^{1/5}D_{27}^{-2}
\notag \\
&&\times f_{b}\frac{\sqrt{\sigma }}{\left( 1+\sigma \right) ^{3/2}}.
\notag
\end{eqnarray}

In Figures 1 to 3, we can find that the dynamic evolutions of the
shocks are only slightly dependent on the magnetar wind
magnetization, but the characteristic frequencies and light curves
of the reverse shock would be
affected by the $\sigma $-value. The other parameters such as $M_{\text{ej}}$%
, $n$ and $L_{0}$ could also play an important role in the
brightness and evolution of the afterglow. In our calculation, we
take the external medium density $n=1$ cm$^{-3}$.

\section{Discussions and Conclusions}

\label{sec:discuss}

Long-lasting energy injection from a post-merger millisecond
magnetar wind has been proposed to drive multi-component
electromagnetic counterparts to gravitational wave bursts
\citep{gao13,wang13b,wang15,yu13,metzger14,gao15}. Current models
concentrate on two extreme cases. In the first case, a pure
Poynting flux drives the forward shock into the ambient medium %
\citep{gao13,wu14}. This is the $\sigma \rightarrow \infty $ extreme
case. In the other extreme case, i.e. $\sigma \rightarrow 0$, the
Poynting flux within the magnetosphere of the magnetar is
transformed into kinetic energy of leptons by magnetic reconnection
\citep{wang13b,wang15}. In this paper, we suggest that in most cases
the magnetar wind is more likely to be a mixture of Poynting flux
and ultra-relativistic $e^{\pm }$ pairs.

The comoving magnetic field strength in the emission regions is a
critical parameter for the synchrotron radiation. For Region 2 (the
forward shocked medium), because the ambient medium is usually
unmagnetized, the downstream
magnetic field may be produced by the relativistic two-stream instability %
\citep{medvedev99}. This magnetic field strength is quantified by
the equipartition parameter $\varepsilon _{B,f}$, i.e. the ratio of
magnetic energy density\ to the total energy density behind the
forward shock. The strength of this magnetic field is quite low for
$\varepsilon _{B,f}$ in the range $10^{-4}$-$10^{-2}$, as inferred
by fitting multi-wavelength afterglows of GRBs
\citep{panaitescu02,yost03}. In addition, this magnetic field may be
small-scale and random.

The magnetic field in the reverse-shocked region could be quite
different because its origin depends on the magnetization of the
upstream cold wind. \cite{wang13b} and \cite{wang15} assumed that
the upstream fluid of the reverse shock is unmagnetized. To
calculate synchrotron radiation from Region 3, a random magnetic
field is introduced through the equipartition parameter $\varepsilon
_{B,r}$, which can be written as
\begin{eqnarray}
B_{3,\text{ra}}^{\prime } &=&\sqrt{8\pi \varepsilon
_{B,r}e_{3}}=\sqrt{8\pi \varepsilon _{B,r}\left( 4\Gamma
_{34}+3\right) \left( \Gamma _{34}-1\right) n_{4}^{\prime
}m_{e}c^{2}}
\end{eqnarray}

In this paper, we suggest that the magnetization of the unshocked
wind could be so large that the magnetic field in the
reverse-shocked region is amplified by compressing the upstream
magnetic field. The strength of this magnetic field is dependent on
the magnetization parameter $\sigma $. Because the field of
shock-compressed upstream magnetic field is globally structured, the
reverse shock emission may have a high polarization.

In all dynamical cases, when $t>T_{\text{dec}}$ the dynamic
evolution of the blast wave obeys the self-similar solution of
\cite{blandford76}. The evolution of the blast wave depends on the
total energy $E$ and the external medium density $n$ and is almost
independent of the magnetar wind magnetization when
$t>T_{\text{dec}}$\ (See Figures 1-3). Such a result has
been found in the previous works %
\citep{emmering87,bucciantini02,mimica09,mimica10,mao10}.

The reverse shock emission is influenced by the degree of
magnetization. On the one hand, a weak magnetic field (in the
low-$\sigma $ regime case) would suppress the synchrotron radiation.
The formation of a reverse shock always takes place in a weakly
magnetized wind as long as the shocked wind could move
supersonically with respect to the unshocked wind. On the other
hand, a strong magnetic field (in the high-$\sigma $ regime case)
would reduce the energy fraction of the shock-accelerated electrons.
As shown in Figure 4, the peak flux of reverse shock emission is a
function of $\sigma $. We find that $\sigma \sim 0.3$ leads to the
strongest reverse shock emission.

It is important to note the difference between the magnetized wind
and the magnetized ejecta. For the later case, in which the
composition is baryons, \cite{mimica10} showed that the RS emission
peaks for $\sigma \sim 0.1$\ (their Figures 3 and 6). It is also
shown that for $\sigma \sim 1$\ the RS
emission should disappear, while \cite{giannios08} shows that even for $%
\sigma \sim 0.3$\ the RS may vanish for some combinations of the
wind luminosity and the external medium density. The disappearance
of RS in this case is linked to the low relative Lorentz factor
between Region 3 and Region 4, as can be seen in the lower panel of
Figure 3 in \cite{mimica09}. For the magnetized magnetar wind we
focus on in this paper, for typical parameters of the\ ejecta and
the\ magnetar, e.g. $\Gamma _{w}>10^{4}$,\ the maximum Lorentz
factor of the blast wave is $\Gamma \sim 10$, which means that
$\Gamma _{34}\approx \Gamma _{w}/2\Gamma \gg 1$ is always satisfied.
This is also true for the prerequisite $\Gamma _{34}\gg
f_{a}f_{b}f_{c}$\ as
determined by \cite{zhang05} because $f_{a}<1$, $f_{b}<1$, and $f_{c}\sim 10$%
. A strong reverse shock would therefore exist even in the high
magnetization regime.

Figures 1-3 shows that the light curves for different values of the
magnetization parameter are quite similar. Consequently it is not
easy to measure magnetization directly from light curves.\ To
measure the magnetization, polarization should be measured. The
successful polarization measurement of the optical afterglow
emission of GRB090102 obtain a linear polarization degree about 10\%
\citep{steele09}. It indicates large scale magnetic fields in the
emission region \citep{steele09,mimica10,harrison13}.

\begin{acknowledgements}
 This work is supported by the National Basic Research Program (``973" Program)
 of China under Grant No. 2014CB845800 and the
 National Natural Science Foundation of China (grant Nos. 11573014 and 11473008).
\end{acknowledgements}

\clearpage
\begin{figure}[tbp]
\begin{center}
\includegraphics[width=0.5\textwidth]{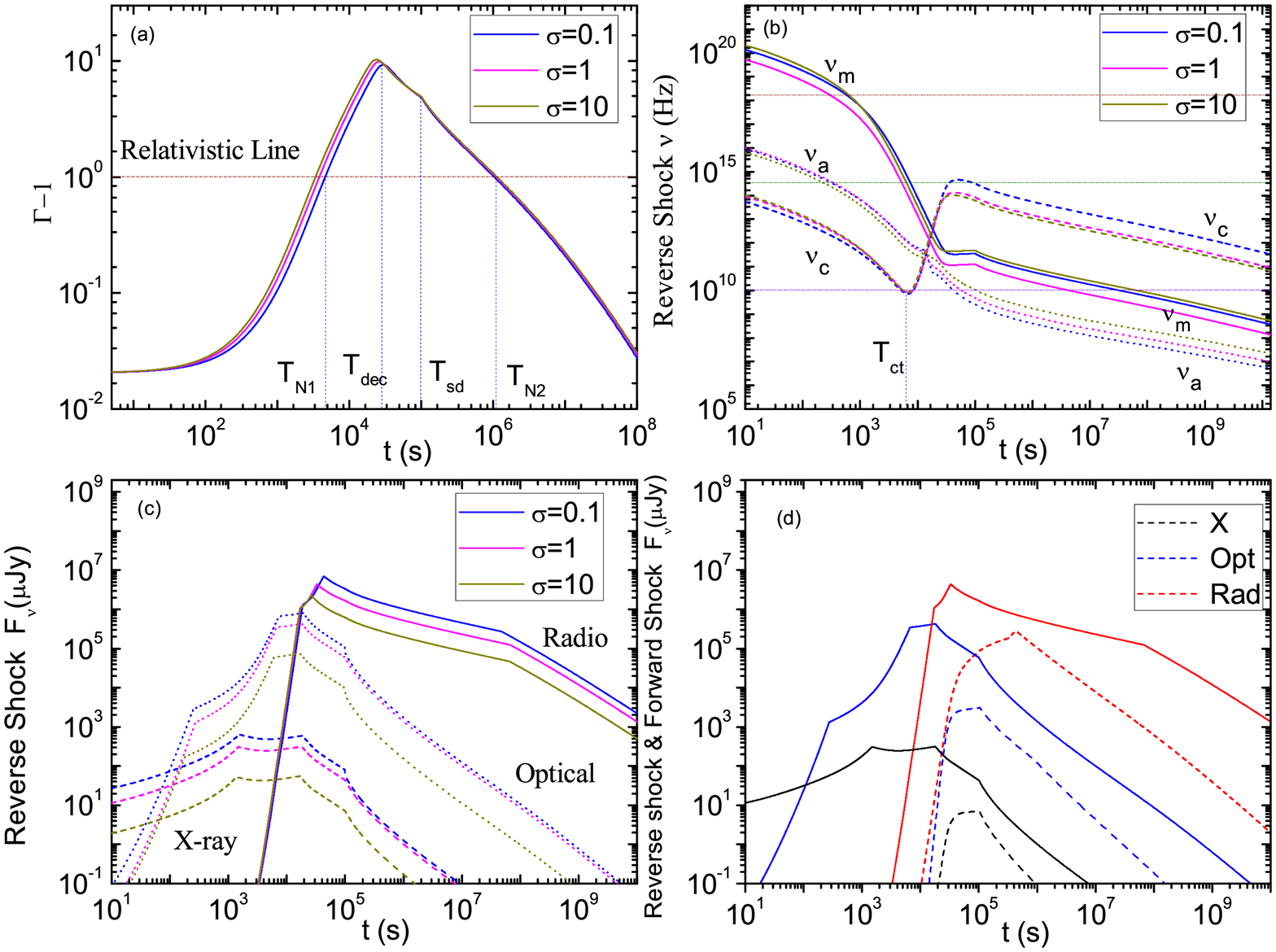}
\end{center}
\caption{Calculation results for Case I: $L_{\mathrm{0}}=10^{47}$
erg s$^{-1} $, $M_{\mathrm{ej}}=10^{-4}M_{\odot }$, $\protect\xi
=0.5$, $p=2.3$. (a) The
evolution of the Lorentz factor of the blast wave with different values of $%
\protect\sigma $ as labeled. (b) The break frequencies $\protect\nu
_{a}$ (dotted), $\protect\nu _{m}$ (solid) and $\protect\nu _{c}$
(dashed) of the reverse shock emission. The three dashed to dotted
lines mark the X-ray, optical ($R$) and radio (10 GHz) bands,
respectively.(c) Light curves of the reverse shock in X-ray, optical
($R$) and radio (10 GHz) bands. (d) The
solid lines represent the reverse shock light curves for $\protect\sigma %
=0.1,$ and the dashed lines are light curves for the forward shock.
In
panels (a), (b) and (c), blue, magenta and dark yellow lines represent $%
\protect\sigma =0.1,1$ and $10,$ respectively. In panel (d), we can
obtain that in Case I the reverse shock emission is stronger than
forward shock emission. It can be seen that the X-ray flux lasts for
several tens seconds
while the optical flux maintain a high level until the spin down time $T_{%
\text{sd}}$, and then the radio emission becomes dominant. }
\label{fig for case I}
\end{figure}

\begin{figure}[tbp]
\begin{center}
\includegraphics[width=0.5\textwidth]{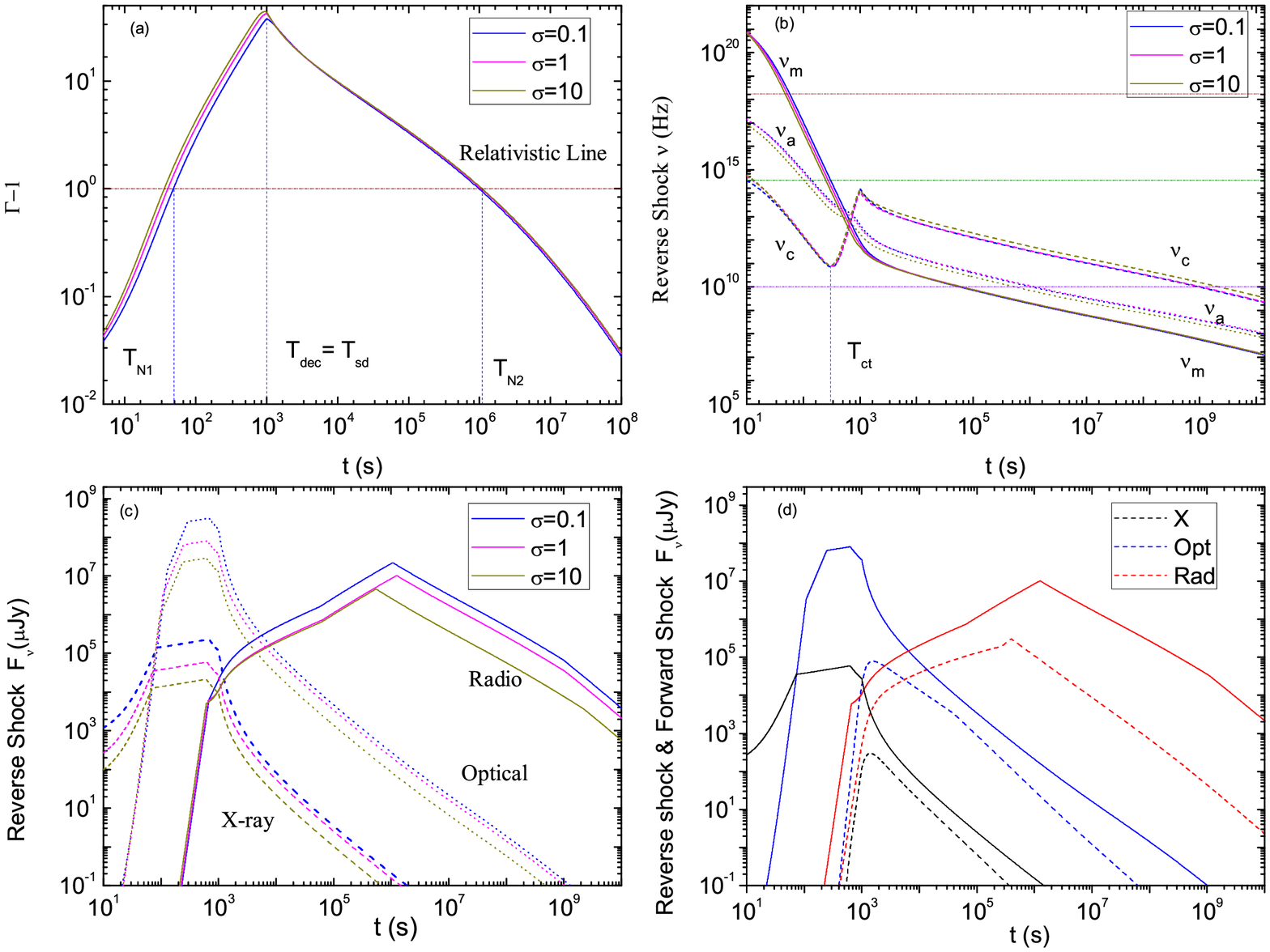}
\end{center}
\caption{Calculation results for Case II: $L_{\mathrm{0}}=10^{49}$ erg s$%
^{-1}$, $M_{\mathrm{ej}}=10^{-4}M_{\odot }$. Descriptions of panels
are the same as in Figure 1. In panel (a), the blast wave
decelerates immediately after linearly accelerating. In panes (c)
and (d), due to more energy injection to the blast wave than Case I,
both reverse shock and forward shock emission is stronger than that
in Case I. } \label{fig for Case II}
\end{figure}
\begin{figure}[tbp]
\centering
\includegraphics[width=0.5\textwidth]{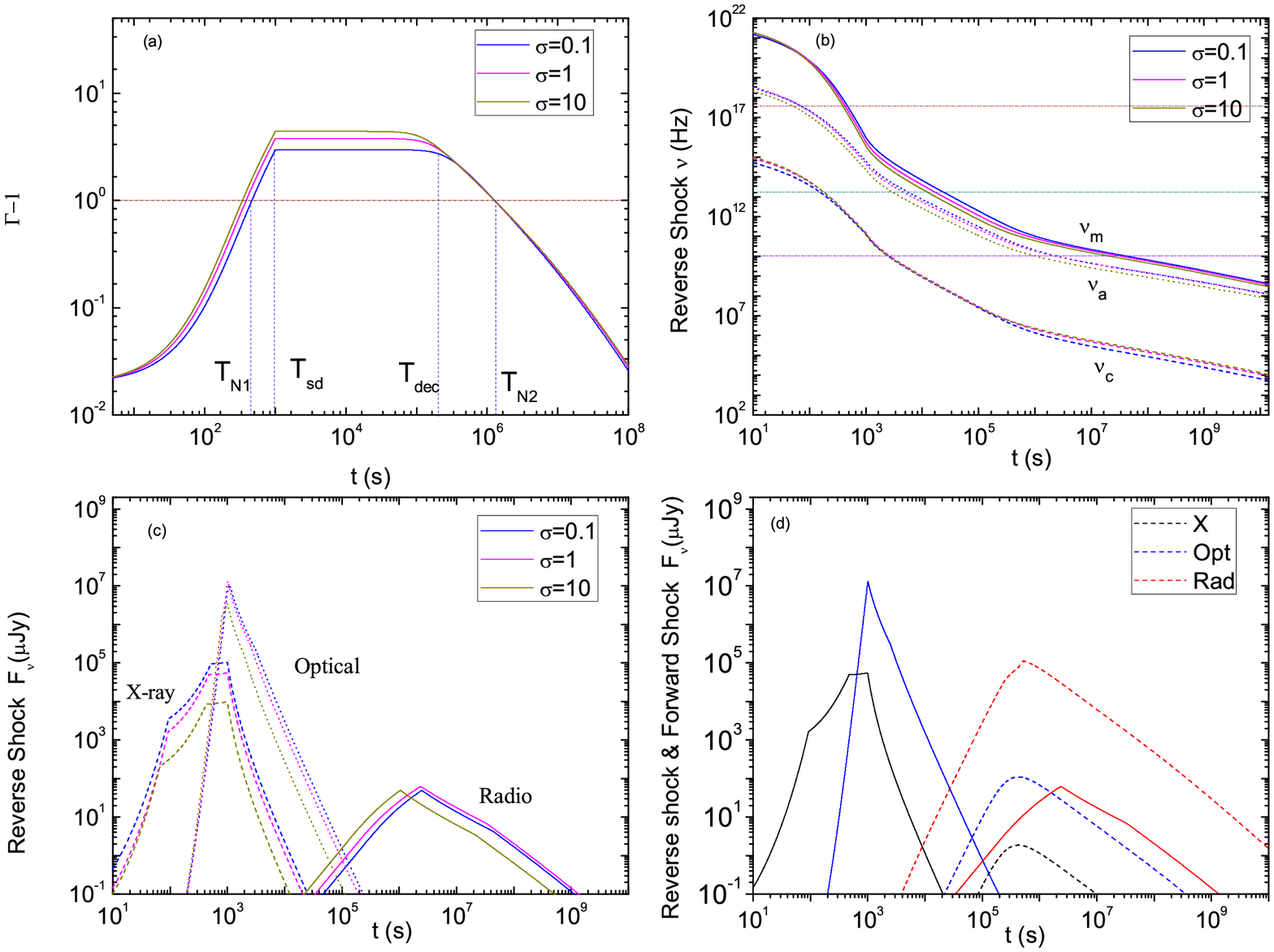}
\caption{Calculation results for Case III: $L_{\mathrm{0}}=10^{49}$ erg s$%
^{-1}$, $M_{\mathrm{ej}}=10^{-3}M_{\odot }$. Descriptions of panels
are the same as in Figure 1. In panel (a), the dynamic of blast wave
shows a coasting phase between $T_{\text{sd}}$ and $T_{\text{dec}}$.
In panes (b),
the break frequencies $\protect\nu _{a}$ , $\protect\nu _{m}$ and $\protect%
\nu _{c}$ without any intersections each other.} \label{fig for case
III}
\end{figure}

\begin{figure}[tbp]
\centering\includegraphics[width=0.5\textwidth,]{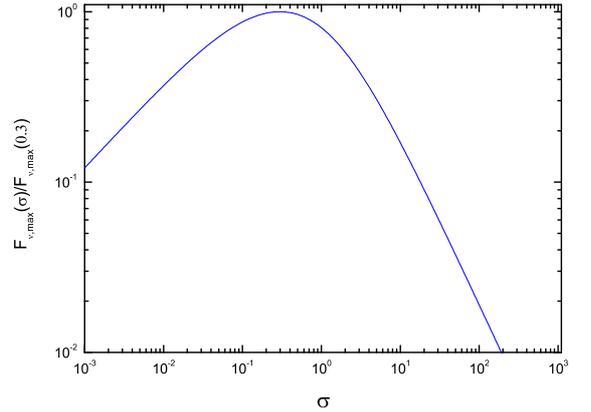}
\caption{Peak flux of the reverse shock emission as a function of $\protect%
\sigma $. It can be seen that the $\protect\sigma \sim 0.3$ leads to
the strongest reverse shock emission. The emission of magnetized
lepton-dominated wind as a function of magnetization is quite
different from that of magnetized baryon-dominated ejecta. See the
main text for more detail.} \label{fig fvmax}
\end{figure}

\clearpage

\end{document}